# Emergence of Chaos in Magnetic-Field-Driven Skyrmions

Gyuyoung Park and Sang-Koog Kim[a]

*National Creative Research Initiative Center for Spin Dynamics and Spin-Wave Devices, Nanospinics Laboratory, Research Institute of Advanced Materials, Department of Materials Science and Engineering, Seoul National University, Seoul 151-744, Republic of Korea*

We explore magnetic-field-driven chaos in magnetic skyrmions. Oscillating magnetic fields induce nonlinear dynamics in skyrmions, arising from the coupling of the secondary gyrotropic mode with a non-uniform, breathing-like mode. Through micromagnetic simulations, we observe complex patterns of hypotrochoidal motion in the orbital trajectories of the skyrmions, which are interpreted using bifurcation diagrams and local Lyapunov exponents. Our findings demonstrate that different nonlinear behaviors of skyrmions emerge at distinct temporal stages, depending on the nonlinear dynamic parameters. Investigating the abundant dynamic patterns of skyrmions during the emergence of chaos not only enhances device reliability but also provides useful guidelines for establishing chaos computing based on skyrmion dynamics.

a) Author to whom all correspondence should be addressed; electronic mail: sangkoog@snu.ac.kr

**Introduction**

Chaos presents a fundamental phenomenon in nonlinear dynamics, observed in various systems, both classical and quantum mechanical oscillators. Nonlinear dynamical systems can exhibit diverse behaviors, including fixed points, periodic and quasiperiodic, and chaotic motions. This transition from the regular nonlinear behavior to chaos is a typical process on the route to chaos. Importantly, these behaviors can manifest at different temporal stages within a single system[1]. For instance, a system may initially exhibit periodic motion before transitioning to chaos, or it may start with quasiperiodic motion and gradually become chaotic over time.

In the field of spintronics, topologically protected spin textures like skyrmions have garnered significant attention due to their potential applicability in low-power-driven high-density data storage[2,3], and ultrafast information processing, thanks to their topological stability at the nano-scales[4]. Therefore, delving into the nonlinearity within skyrmion dynamics and any potential chaotic effects is crucial for ensuring reliable device utilization. While chaotic dynamics of other topological spin textures, such as current-driven magnetic vortices[5,6] and antiferromagnetic bimerons[7], have been studied, reports on chaotic dynamics of skyrmions are relatively rare, even though their nonlinear dynamics have been examined in previous studies[8]. Further research is needed to fill this gap and explore chaotic dynamics in magnetic skyrmions.

In this letter, we begin by investigating the field-driven nonlinear dynamics of magnetic skyrmions, which exhibit complex hypotrochoidal trajectories. Subsequently, we map the various regimes of the skyrmion's nonlinear behavior over time, building a comprehensive nonlinear dynamic phase map. This map illustrates the transition from regular to chaotic regime along temporal space versus the amplitude of the oscillating magnetic fields. Additionally, by exploring and harnessing the chaotic behavior of skyrmions, we may pave the way for novel applications in chaos-based computing and signal processing. Therefore, our findings provide valuable insights for designing future skyrmion-based technologies with enhanced functionalities and reliability.

**Methods**

The model employed in the present study involves a Néel-type skyrmion formed in a circular nano-disk with a radius ($R_{disk}$) of 30nm and a thickness of 0.6 nm, as shown in its ground state in Fig. 1(a). To reduce the magnetostatic energy at the boundary of the disk, spins at the edge are tilted toward the center of the skyrmion. To numerically calculate the dynamic motion of the skyrmion, we used the MuMax³ code[9], which utilizes the Landau-Lifshitz-Gilbert (LLG) equation: $\partial \mathbf{M}/\partial t = -\gamma \mathbf{M} \times \mathbf{H}_{eff} + (\alpha_G/M_s)\mathbf{M} \times \partial \mathbf{M}/\partial t$, where $\gamma$ is the gyromagnetic ratio, $\alpha_G$ is the Gilbert damping constant, and $\mathbf{H}_{eff}$ is the effective field given as $\mathbf{H}_{eff} = -(1/\mu_0)\partial E_{tot}/\partial \mathbf{M}$. The total energy, $E_{tot}$, includes the magnetostatic, magnetocrystalline anisotropy, exchange, and intrinsic Dzyaloshinskii-Moriya Interaction (DMI), and Zeeman energies. For the material, we assumed Co interfaced with Pt and used the following parameters: saturation magnetization $M_s$ = 0.58×10⁶ A/m, exchange constant $A_{ex}$ = 1.5×10¹¹ J/m, perpendicular anisotropic constant $K_u$ = 4×10⁵ J/m³, Gilbert damping constant $\alpha_G$ = 0.01, and interfacial DMI constant $D_{int}$ = 3 mJ/m². The cell size was set to 0.6×0.6×0.6 nm³.

With the aforementioned conditions, we intentionally formed a Néel-type skyrmion with a core polarization of -1 (spin down) and allowed it to relax to reach its ground state. In the ground state, the skyrmion exhibits a circular domain wall represented by the gray line and inset circle, indicating a radius ($R_{sky}$)[10] of about 12.5 nm. We applied a linearly polarized harmonic field, $H_{AC} = \alpha sin(2\pi f_+ t)$, uniformly over the entire disk, where $\alpha = A/A_s$[11] and $A_s$ (= 2730 Oe) represents the static field used to annihilate the skyrmion. We gradually increased $\alpha$ from 0 to 0.30 in increments of $5 \times 10^{-4}$. By considering the skyrmion's inertial mass $\mathcal{M}$[12] and using Thiele's equation of motion[13],

$$-\mathcal{M}\ddot{\boldsymbol{R}} + \boldsymbol{G} \times \dot{\boldsymbol{R}} - K\boldsymbol{R} = 0. \qquad (1)$$

We obtained the angular frequencies of two gyrotropic modes as follows:

$$w_{\pm} = -G/2\mathcal{M} \pm \sqrt{(G/2\mathcal{M})^2 + (K/\mathcal{M})}. \qquad (2)$$

The higher (secondary) gyrotropic mode, denoted as $f_+$ ($f_+ = w_+/2\pi \approx 21.37$ GHz, cf. $f_- \approx$ 2.19 GHz at zero external field), represents a counter-clockwise rotation in this core polarity of spin down. Here, $G$ is the gyrocoupling constant, $K$ is the spring constant, and $\boldsymbol{R} = (X, Y)$ represents the guiding center with $X = \int xq dxdy / \int q dxdy$ and $Y = \int yq dxdy / \int q dxdy$, where $q = (1/4\pi)\boldsymbol{m} \cdot (\partial_x \boldsymbol{m} \times \partial_y \boldsymbol{m})$ is the topological charge density[14].

Unlike the lower or fundamental gyrotropic mode, $w_-$, where the entire skyrmion moves as a rigid body, the higher or secondary gyrotropic mode, $w_+$, exhibits a unique behavior (see supplemental material). In the $w_+$ mode, the core and peripheral spins gyrate in the same direction, but they are out-of-phase when the skyrmion is excited[15]. This leads to a coupling between the in-plane gyrotropic mode and a non-uniform out-of-plane breathing[16]-like mode. Consequently, during the application of a driving force, the skyrmion undergoes a deformation that breaks its rotational symmetry. When the topological soliton experiences deformation, its moment of inertia and spring constant are altered, giving rise to the soliton's nonlinear dynamics[17,18]. This deformation-induced nonlinear dynamics exclusively occurs when the skyrmion is driven by the higher gyrotropic mode. In Fig. 1(b), we illustrated the deformed skyrmion during its gyration. The Gray circle shows the distorted contour of the skyrmion's domain wall, which assumes an irregular ellipse-like shape due to the broken rotational symmetry of the breathing-like mode being coupled with its gyration mode. A profile of the individual spins' orientations across the center is shown below the skyrmion illustration. Compared to the ground state, the number of black spins inside the skyrmion's domain wall increases from 5 to 7. Small black circles just above the black spins denote the position of the guiding center at that moment. The guiding center follows an orbital motion during the gyration, resembling a hypotrochoid, which belongs to the family of curves known as roulettes.

**Results**

The hypotrochoid is traced by a point attached to a circle with radius $b$, rolling inside a fixed circle with radius $a$, where the point is a distance $d$ from the interior circle's center. When the higher gyrotropic mode is excited, its orbital motion represented by the guiding center inevitably follows one of the hypotrochoids[12,19], which can be parametrized as follows:

$$z(t) = r_1 e^{iw_1 t} + r_2 e^{-iw_2 t}, \quad (3)$$

where the geometric parameters are related as follows: $d = r_2$, $b = (w_1/w_2) r_1$, and $a = ((w_1 + w_2)/w_2) r_1$. The Thiele's equation (Eq. 1) can be modified with the linearly-polarized oscillating magnetic field as:

$$-\mathcal{M}\ddot{\boldsymbol{R}} + \boldsymbol{G} \times \dot{\boldsymbol{R}} - K\boldsymbol{R} + \mu(\hat{\boldsymbol{z}} \times \boldsymbol{H}) = 0, \quad (4)$$

with $\boldsymbol{H} = (0,\ \mu\alpha H_s sin(w_+ t),\ 0)$, and the solution of Eq. (4) is,

$$R(t) = R_- e^{iw_- t} + (R_+ + \frac{h}{2})e^{iw_+ t} - \frac{h}{2} e^{-iw_+ t} \quad (5)$$

with $h = \mu\alpha H_s \sin^2 w_+ t/(\mathcal{M} w_+^2 + K)$. Eq. (5) takes a form similar to that of the hypotrochoids, Eq. (3). Note that the two gyrotropic modes $w_\pm$ have opposite signs, indicating the opposite rotation sense, with $R_- = -R_+ = (-w_- + w_+)h/w_+$. Therefore, the hypotrochoidal parameters for the skyrmion are given as $r_1 = (-w_- + w_+)h/w_+$ and $r_2 = (R_+ + \frac{h}{2}) = (2w_- - w_+)h/2w_+$. So far, the equation of motion has been linear. However, it becomes nonlinear as the eigenfrequencies, denoted by Eq. 2, vary during the motion. Due to the skyrmion's deformation under oscillation causes changes in $\mathcal{M}$ and $K$, where $\mathcal{M} = \zeta \bar{r}$ and $K = \xi \bar{r}^2$ [19], with $\zeta$ and $\xi$ being deformation-induced coefficients, and $r$ representing the local radial distance of the closed domain wall. Furthermore, $r$ varies significantly for irregularly shaped skyrmions [20].

The shape of the hypotrochoid can be defined by two parameters, namely, the number of cusps, $\nu = a/b = (w_1 + w_2)/w_1$, and the ratio of the distance to the smaller radius, $\varepsilon = d/b = r_2 w_2 / r_1 w_1$. $\varepsilon$ determines the type of the hypotrochoid. For example, in either case of ($w_1 > w_2$ and $\varepsilon < 1$) or ($w_1 < w_2$ and $\varepsilon > 1$), the curve is a prolate hypotrochoid. Conversely, if the opposite conditions are met, the curve is a curtate hypotrochoid. When $\varepsilon = 1$, the curve is a hypocycloid.

Figure 2(a) illustrates the representative shapes of the orbital trajectories for the indicated values of ($\nu$, $\varepsilon$), which are also labeled by ① ~ ⑫ in Fig. 2(b) for four different values of $\alpha$ (i.e., $\alpha = 0.0245, 0.0365, 0.126,$ and $0.1485$). These $\alpha$ values fall within distinct transition spaces between different nonlinear behaviors. The parameter $\nu$ was estimated from the Fast Fourier Transform of the trajectories over time, while $\varepsilon$ was derived from a geometrical analysis of $\mathbf{R}$ ($X$, $Y$) without any fitting. The number of cusps in each orbital geometry corresponds well with each $\nu$ value. Meanwhile, when $\varepsilon = 1$, the hypotrochoid becomes a hypocycloid, as seen in cases ④ and ⑦. In this skyrmion system, $w_1$ ($=w_-$) is always smaller than $w_2$ ($=w_+$). Therefore, trajectories with $\varepsilon > 1$ exhibit prolate hypotrochoids, as observed in cases ③, ⑥, and ⑨, while trajectories with $\varepsilon < 1$ display curtate hypotrochoids, as in cases of ⑧, ⑩, ⑪, and ⑫.

Figure 2(b) depicts the hypotrochoidal parameters of $\nu$ (height) and $\varepsilon$ (color) as functions of evolution time ($\tau$) and reduced field amplitude $\alpha$ (0 ~ 0.157) for skyrmion motions excited by oscillating magnetic fields. It is important to note that beyond $\alpha = 0.157$, the skyrmion breaks down within the observation time from 0 to 534 $\tau$, where $\tau$ represents one period of the modulation frequency and is equivalent to the inverse of the frequency of the higher gyrotropic mode, $1/f_+$. In the plots of ($\nu$, $\varepsilon$), two distinct regions (black color) are separated by the maximum $\varepsilon$ values (yellow color) along given alpha values. When the condition of $\tau$ and $\alpha$ below the yellow line, the skyrmion motions show fixed points or periodic motions, while they exhibit quasiperiodic and chaotic motion on the condition above the yellow line. Furthermore, there is a transition from prolate-type to curtate-type after the maximum $\varepsilon$ line.

Next, we turn our attention to various nonlinear regimes related to the nonlinear parameters. Analyzing a single variable, such as the orbital radius $\rho = |\mathbf{R}|/R_{sky}$, offers a highly effective method to classify different types of nonlinear behaviors depending on the nonlinear parameters. In our analysis, we identified six critical points: $\alpha = 0.0245$ for the shift from a fixed point to periodic motion; $\alpha = 0.0365$ for the transition from periodic to quasiperiodic motion; $\alpha = 0.126$ and $\alpha = 0.1485$ for the gradual transitions to complex quasiperiodicities; $\alpha = 0.152$ for the transition from quasiperiodic motion to chaos; and finally, at $\alpha = 0.1565$, the skyrmion begins to break down during its motion.

For a more qualitative analysis of nonlinear behaviors, we plotted $\rho$ versus time for $\alpha = 0.0245, 0.0365, 0.126, 0.1485$, and $0.152$, as illustrated in Fig. 3(a). When the skyrmion is driven by the higher gyrotropic mode, $\rho$ exhibits two steady-states, $\rho_{\pm}$, at different temporal stages: an initial unstable $\rho_{+}$ state, followed by a later stable $\rho_{-}$ state. In the $\rho_{+}$ state, the skyrmion is underdamped, while in the $\rho_{-}$ state, it becomes overdamped due to the changes in $K$. For $\alpha < 0.0245$, the skyrmion doesn't reach the $\rho_{-}$ state, hindered by an energy barrier that primarily originates from the damping force. This regime is a fixed point where the guiding center asymptotically approaches the $\rho_{+}$ state after extensive gyration. As $\alpha$ surpasses 0.0245, $\rho$ increases, and its $\rho_s$ oscillates around the $\rho_{-}$ state. As $\alpha$ continues to increase, the transition time from $\rho_{+}$ to $\rho_{-}$ decreases. Specifically, at $\alpha = 0.0365$, the $\rho_{-}$ state is achieved after a few hundred periods. In this regime of $\alpha < 0.0365$, $\rho$ exhibits a periodic behavior in its steady-states, whereas it becomes quasiperiodic (or loosely periodic) for $\alpha \geq 0.0365$. In this latter regime, some variations in the signal do not conform to regular periodicities. For $\alpha = 0.126$ and $\alpha = 0.1485$, the quasi-periodicities gradually become much more complex. Finally, at $\alpha = 0.152$, regular periodicities start to diminish, and chaos begins to occur. The signals shown at the top of Fig. 3(a) for $\alpha = 0.126, 0.1485$, and 0.152 allow us to better visualize these periodicities. Unlike the cases of $\alpha = 0.126$ and 0.1485, the case of $\alpha = 0.152$ has no periodicities, indicating a transition from quasiperiodic to chaotic motions. When $\alpha \geq 0.152$, a well-known property of chaos, the sensitive dependence on initial conditions (SDI), arises. For these values of $\alpha$, signals with minor differences in initial conditions can result in substantial variations in the resulting signals.

Figure 3(b) displays the two steady states, $\rho_{\pm}$, versus $\alpha$. The separate $\rho_{\pm}$ values were identified when the variance of the $<\rho>$ values within a moving window reached a minimum.

The different nonlinear regimes are labeled as fixed points ('FP'), periodic ('P') and quasiperiodic ('QP') motion, and chaos ('C'). In the chaotic regime, the concept of SDI was utilized. With SDI, even an extremely small variance in the initial conditions can lead to vastly different outcomes. $\rho_\pm$ values obtained from four different initial conditions (see Supplemental Materials) were averaged. In the breakdown regime, indicated by black dotted boxes for $\alpha > 0.152$, $\rho_\pm$ exhibit large variations. This distribution signifies that even an extremely small difference in initial conditions, such as $\Delta q < 10^{-5}$, can significantly alter the lifetime of the skyrmion within this regime. Furthermore, the $\rho_-$ value generally decreases in the breakdown regime, indicative of high instability in the dynamics of the skyrmion. As $\alpha$ increases beyond $\alpha = 0.2365$, unstable $\rho_+$ states are no longer observed.

A bifurcation diagram can facilitate a comprehensive understanding of potential long-term behaviors and highlight different periodicities that arise with changes in a key bifurcation parameter. In this context, $\rho_s$ represents the particular periodic behavior, while $\alpha$ acts as the bifurcation parameter. For example, Fig. 3(c) presents the bifurcation diagram of $\rho_s$ versus $\alpha$, offering a visual guide to the field-driven skyrmion's quasiperiodic route to chaos. The diagram outlines different stages, previously detailed in Fig. 3(b), and marks each bifurcation point with black-dashed vertical lines. As $\alpha$ increases, both $\rho_-$ and $\rho_+$ bifurcate, but the skyrmion does not oscillate between the $\rho_\pm$ states, back and forth. The existence of a breakdown regime and finite simulation time necessitate the inclusion of SDI in the chaos regime, by plotting the $\rho_s$ values at other initial conditions. Unlike the 'FP' regime, $\rho_s$ increases exponentially in the 'P' region. Through a series of bifurcations in the 'QP' regime, denoted as I, II, and III, the skyrmion enters the chaos regime, where $\rho_s$ can take on every possible value, as evidenced by the scattered data points. This result effectively describes the series of quasiperiodic routes to chaos.

Finally, we evaluated the local Lyapunov exponent (LLE), denoted as $\lambda(t)$, a measure of sensitivity to initial conditions in a dynamical system, evaluated locally in the phase space (i.e., at a specific point or over a short period of time). We constructed the $\lambda(t)$ map from the time trace of the speed, $v$, of the guiding center $\boldsymbol{R}$ ($X$, $Y$) as follows [21]:

$$\lambda(t) = \frac{\delta(t + \Delta t) - \delta(t)}{\Delta t} \qquad (6)$$

and

$$\delta(t) = \frac{1}{N} \sum_{i=1}^{N} \ln\left(\frac{\|V_{i+t} - V_{j+t}\|}{\|V_i - V_j\|}\right) \qquad (7)$$

where $\vec{V_i} = [v_i, v_{i+T}]$ is the reconstructed phase space with time delay $T$ and $\vec{V_j}$ is the nearest neighbor such that $\|\vec{V_i} - \vec{V_j}\|$ is minimized. In the resultant LLE map, with respect to $\alpha$ and time ($\tau$), as shown in Fig. 4, blue regions (where $\lambda < 0$) depict the exponential convergence of trajectories in the phase space, representing stable ordered states. In contrast, red regions (where $\lambda > 0$) indicate the exponential divergence of trajectories in the phase space, signifying chaotic states or ordered but unstable states. White regions (where $\lambda \approx 0$), situated between the red and blue regions, represent a transition space between order and chaos, known as the edge of chaos. A black region at the top of the map denotes the breakdown regime.

The previously classified regimes of nonlinear dynamics in the bifurcation map match well with the LLE map at $\tau \approx 0$. However, the LLE map strongly suggests that the nonlinear behavior changes over time, as evidenced in the map of hypotrochoidal parameters. The first white line on the left corresponds exactly to the maximum $\varepsilon$ line in Fig. 2. This line indicates a transition from unstable to stable states (for $\alpha < 0.152$) or from ordered to chaotic states (for $\alpha > 0.152$). For $0 < \alpha < 0.0365$, the skyrmion maintains its ordered state throughout the oscillation. For $0.0365 \leq \alpha < 0.1$, it begins with a stable limit cycle or torus and displays intermittent unstable ordered states due to repelling trajectories between $\rho_+$ and $\rho_-$ states. For $0.1 \leq \alpha < 0.15$, the skyrmion's dynamic states shift from stable state to unstable. For $0.15 \leq \alpha < 0.17$, the skyrmion shows transient chaos in the initial states. For $0.17 \leq \alpha$, the skyrmion displays chaotic dynamics throughout the oscillation until it breaks down.

**Conclusion**

We investigated the nonlinear dynamic behaviors of a magnetic skyrmion driven by oscillating magnetic fields. The nonlinear mode arises from the coupling of gyrotropic modes with a non-uniform breathing-like mode. The deformation of the skyrmion altered key properties such as its moment of inertia and spring constant, which in turn affected the eigenfrequency of the fundamental gyrotropic mode. This led to changes in the characteristic parameters, including the cusp number and the type of the hypotrochoid, leading to the complex hypotrochoidal motions of the skyrmion over time. By constructing a bifurcation map and a local Lyapunov exponent map, we were able to distinguish the entire range of nonlinear behaviors, spanning from ordered regimes to chaos, across various field amplitudes and temporal stages. This in-depth understanding of the field-strength and time-dependent chaotic dynamics of magnetic skyrmions provides essential insights into different nonlinear dynamic routes to chaos. Furthermore, it can foster the development of innovative computation schemes that leverage the initial condition-sensitive, deterministic chaos dynamics.

**Acknowledgment**

This research was supported by the Basic Science Research Program through the National Research Foundation of Korea (NRF), funded by the Ministry of Science, ICT, and Future Planning (Grant No. NRF-2021R1A2C2013543). The Institute of Engineering Research at Seoul National University provided additional research facilities for this work.

**Figure captions**

Fig. 1. (a) Ground-state magnetization ($m = M/M_S$) configuration of a Néel-type skyrmion confined within a nano-disk. (b) A snapshot of the magnetization configuration of the same skyrmion at an arbitrary moment, excited by a sinusoidal magnetic field ($H_{AC}$) applied along the y-axis. The domain wall of the skyrmion is represented by a gray circle shown in each spin configuration. Inset figures show spin profiles of the gray-shaded boxes. Black wide circles indicate the position of the guiding center at each state.

Fig. 2. (a) Orbital trajectories of skyrmion motions calculated using the guiding center over a time period of $t_i \sim t_i + 1/f_-$ at indicated values of ($\nu$, $\varepsilon$) (see ① ~ ⑫), (b) Calculations of $\nu$ and $\varepsilon$ on the plane of evolution time ($\tau$) and reduced field amplitude ($\alpha$). The $\varepsilon$ value is represented by the color shown in the color bar on the right. White dotted lines correspond to $\alpha = 0.0245, 0.0365, 0.126,$ and $0.1485$.

Fig. 3. (a) The orbital radius $\rho$ versus time calculated using of the guiding center for selected field amplitudes ($\alpha = 0.0245, 0.0365, 0.126, 0.1485,$ and $0.152$). (b) Two steady-states $\rho_\pm$ as a function of $\alpha$. (c) The bifurcation diagram of $\rho_s$ as a function of $\alpha$. 'F', 'P', 'QP', and 'C' symbolize the regime of distinct nonlinear behaviors of fixed point, periodic, quasiperiodic, and chaotic motion.

Fig. 4. Calculation of the local Lyapunov exponent (LLE) as a function of time ($\tau$) and reduced field amplitude, $\alpha$. The color bar indicates the values of the LLE.

Figure 1 (one column)

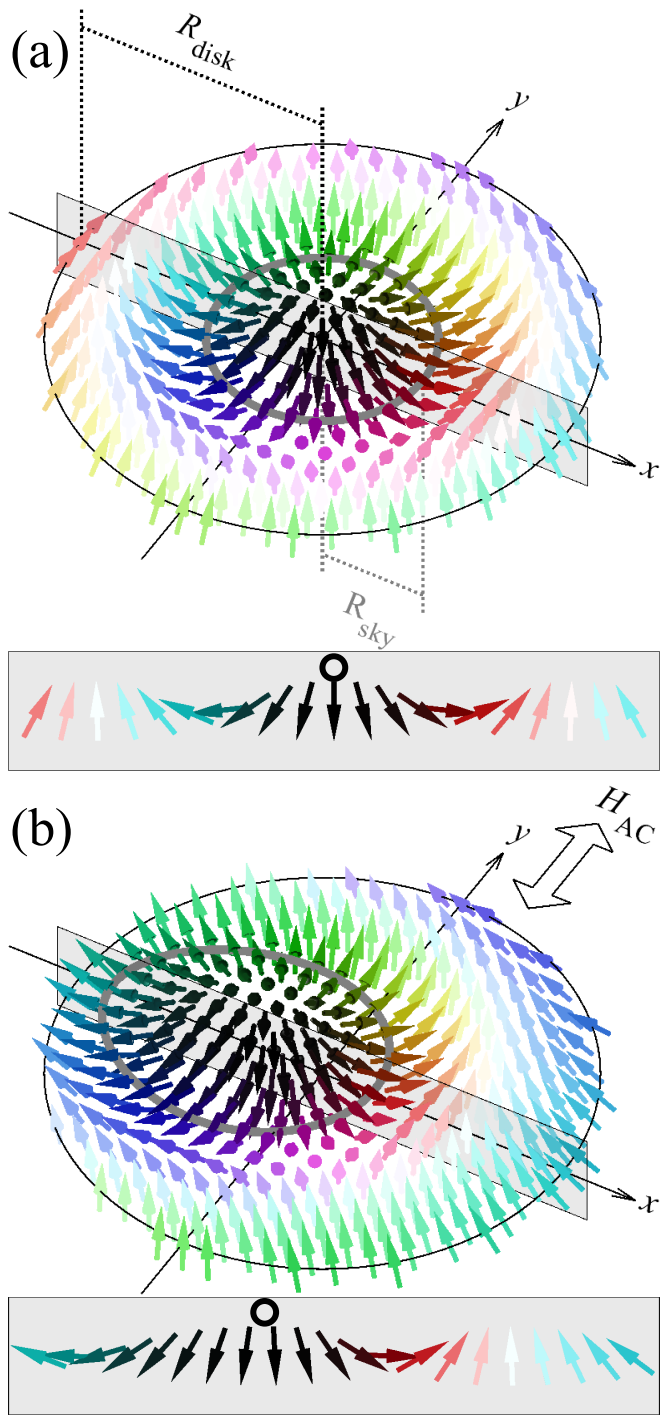

Figure 2 (one column)

(a)
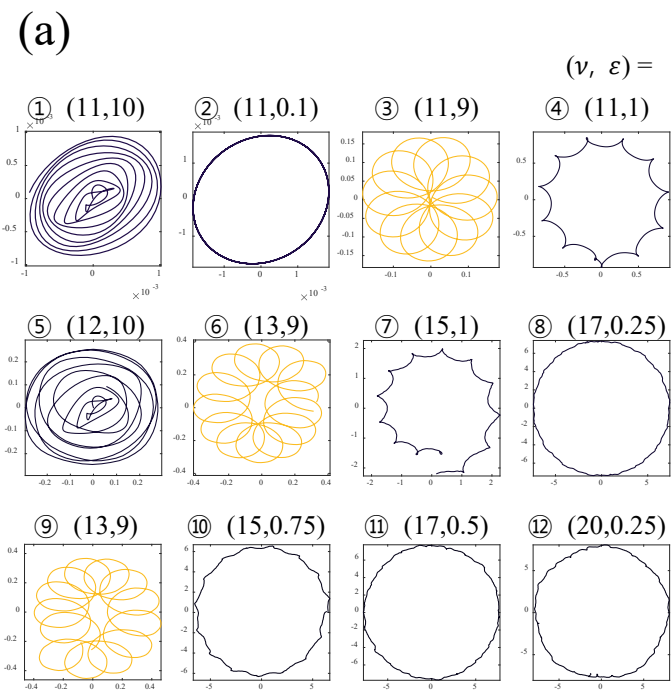

(b)
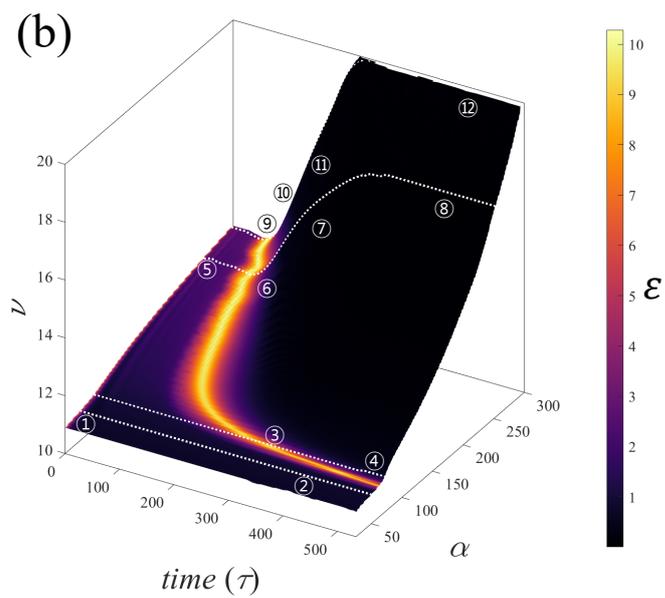

Figure 3 (one column)

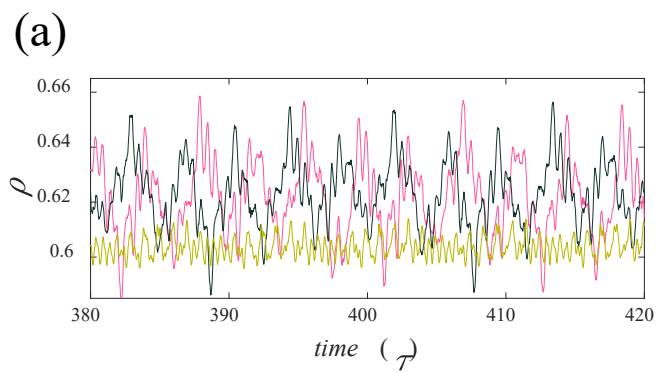
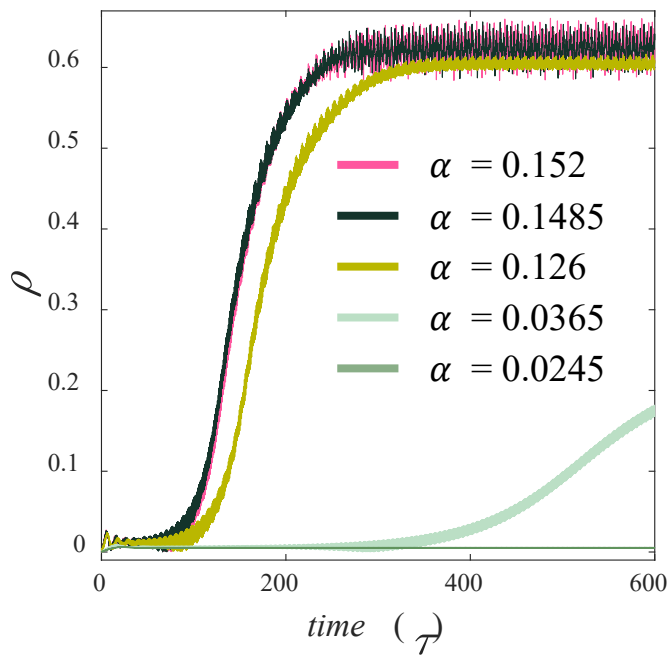
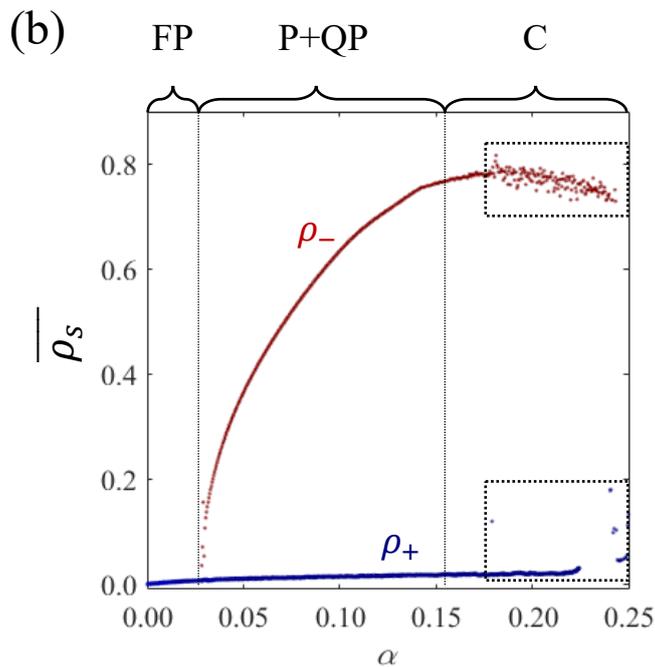
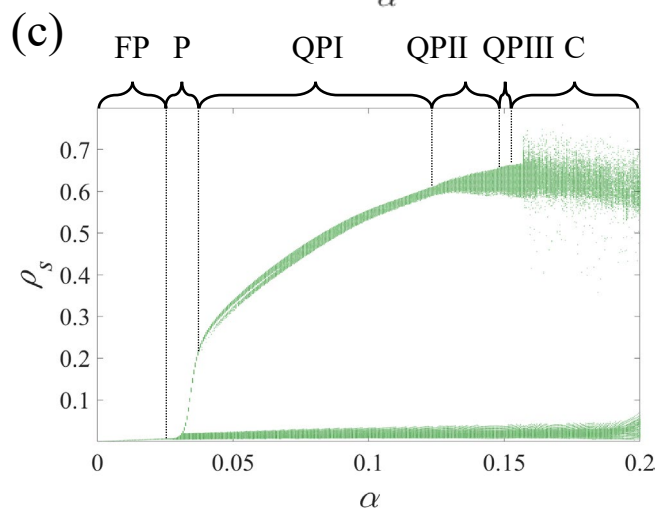

Figure 4 (one column)

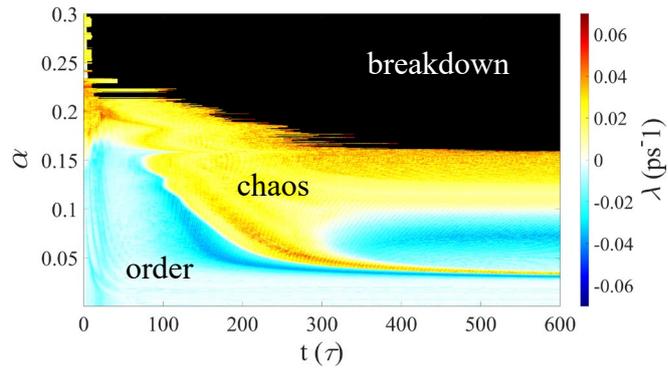

Supplemental Materials

# Emergence of Chaos in Magnetic-Field-Driven Skyrmions

Gyuyoung Park and Sang-Koog Kim[a)]

*National Creative Research Initiative Center for Spin Dynamics and Spin-Wave Devices, Nanospinics Laboratory, Research Institute of Advanced Materials, Department of Materials Science and Engineering, Seoul National University, Seoul 151-744, Republic of Korea*

a) Author to whom all correspondence should be addressed; electronic mail: sangkoog@snu.ac.kr

**Appendix A: Linearity and nonlinearity of the in-plane modes**

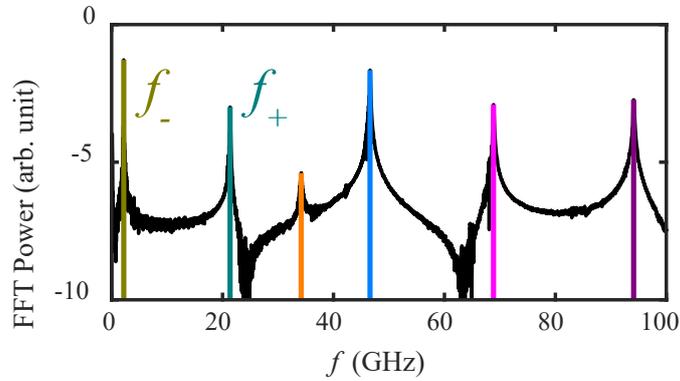

**Figure S1**. The FFT spectrum for the in-plane excitations of the skyrmion driven by a sinc-function field along the *y*-axis.

To obtain a spectrum of in-plane excitation modes for the magnetic field-driven skyrmion, we initially applied a sinc-function magnetic field given by $H_{sinc}(t) = H_0 sinc(2\pi f_H t)$ over the entire nanodisk where $\boldsymbol{H}(t) = (0, H_{sinc}(t), 0)$. Here, $H_0$ is 10 Oe and $f_H$ is 100 GHz. The modes' spectrum, depicted in Fig. S1, was determined using a Fast-Fourier-Transform (FFT) of $m_y$ components of the spins. Two gyrotropic modes appear at 2.19 GHz and 21.37 GHz, while other in-plane excitations, known as azimuthal modes, manifest in higher frequency domains.

To assess the linearity of each distinct mode, we applied a linearly polarized harmonic field corresponding to each mode's frequency along the y-axis. We then measured the $\ddot{X}$ of the guiding center at the initial phase of the field application while varying the field amplitude, $\alpha$, as shown in Fig. S2(a). Notably, $\ddot{X}$ only increases nonlinearly at the higher gyrotropic modes. See the supplemental video to check the raw data. Fig. S2(b) displays the FFT results of the $Y$ component of the guiding center when the skyrmion is driven by a linearly polarized harmonic field, aligned to the higher gyrotropic mode's frequency, $A\sin(2\pi \sin f_+ t)$, with $A$ being 10 Oe, along the y-axis. All six modes were simultaneously excited, even though the harmonic frequency was $f_+$[1]. In contrast, when driven by the $f_-$ frequency, modes other than the lower gyrotropic mode weren't stimulated. Yet, this simultaneous excitation of different in-plane modes doesn't account for the nonlinearity, as their coupling represents a straightforward linear combination of the modes. Azimuthal modes also exhibited similar FFT spectra. Lastly, Fig. S2(c) illustrates the spatial distribution of the FFT phase for the $f_+$ frequency. Both the core and peripheral spins gyrate in a counter-clockwise direction but are in antiphase, with an azimuthal number of 0.

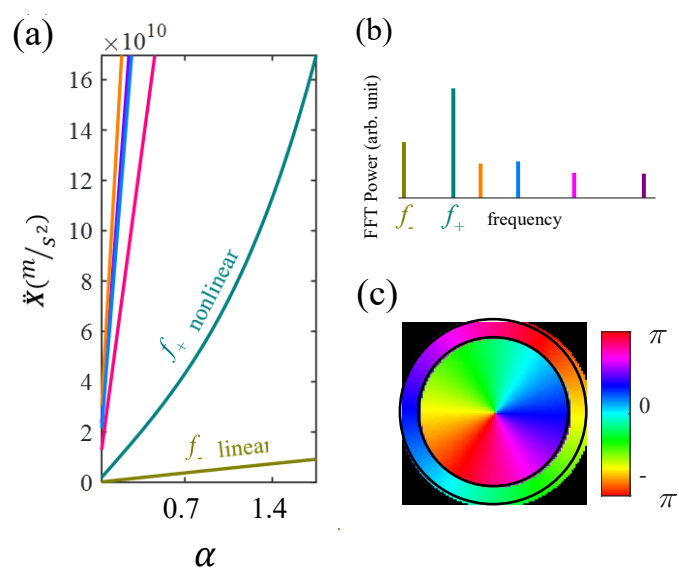

**Figure S2**. (a) Linearity and non-linearity of the in-plane excitation modes in relation to the amplitude ($\alpha$) of the applied oscillating magnetic field. Each graph color corresponds to the colors indicated in Fig. S1. (b) Abbreviated FFT spectrum of the Y component of the skyrmion's guiding center when driven by the $f_+$ mode. (c) FFT phase distribution of the $f_+$ frequency.

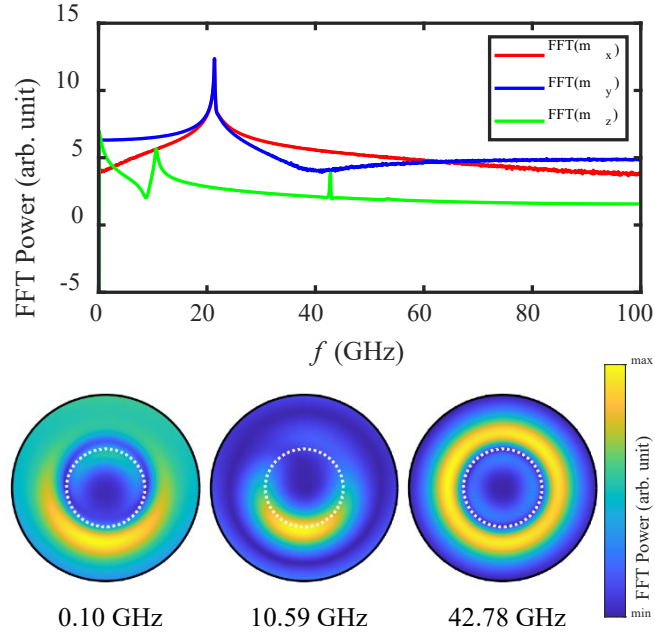

0.10 GHz     10.59 GHz     42.78 GHz

**Figure S3.** The FFT spectrum for the $m_x$, $m_y$, and $m_z$ components of the spins is shown when the skyrmion is driven by an oscillating magnetic field at the frequency of the higher gyrotropic mode. The inset displays the spatial distributions of the FFT power for three distinct out-of-plane modes. White dashed circles denote $R_{sky}$.

Figure S3 presents the mode spectra for both in-plane and out-of-plane directions when the skyrmion is excited by a higher gyrotropic mode. At the frequency of $2f_+$, a uniform breathing mode larger than $R_{sky}$ is excited. At frequencies lower than the $f_+$ mode, two non-uniform out-of-plane modes couple to the $f_+$ mode. This is primarily because the two chiral edge spin wave modes possess different rotational senses[2]. The rotational symmetry of the FFT power distribution is disrupted, suggesting that these modes induce deformation in the skyrmion. A simple, non-uniform breathing mode could distort the skyrmion into an ellipse. However, this non-uniform breathing-like mode is intertwined with the gyrotropic mode. As a result, this combined mode contorts the skyrmion into a shape devoid of symmetries. The amorphous form of the deformed skyrmion complicates analytical approaches.

**Appendix B: Skyrmion breakdown**

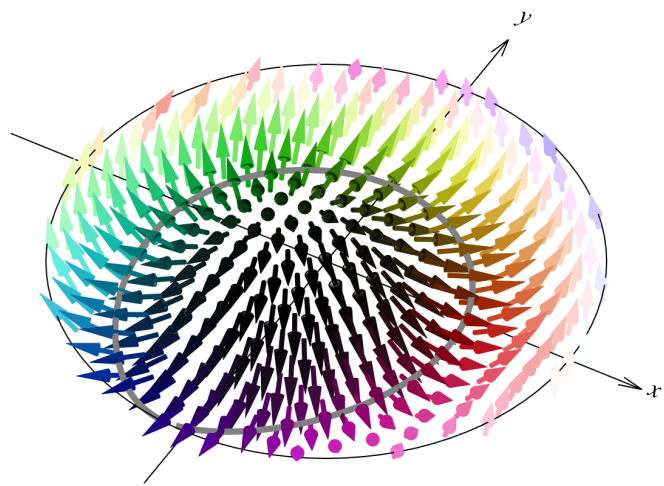

**Figure S4.** The magnetization ($\boldsymbol{m} = \boldsymbol{M}/M_S$) configuration of the skyrmion breaks down. The closed domain wall opens at the bottom of the figure, as indicated by the gray line.

*Breakdown* – A skyrmion, when driven, can break down if the driving current density surpasses a critical value, either due to overpowering the damping force or inducing significant deformation in the skyrmion[3,4]. This breakdown signifies the skyrmion losing its topological charge or its closed domain wall disappearing. Practically, establishing a numerical boundary for the topological charge—where it's still considered a skyrmion—can be challenging. Consequently, we define the skyrmion's breakdown as the point when its closed-loop domain wall opens, primarily due to the loss of its topological structure when approaching the disk boundary (Fig. S4).

*Static breakdown field* - The static breakdown field ($A_s$) is characterized as a linear, static magnetic field uniformly applied across the skyrmion, leading to its disintegration upon sustained application. In our system model, breakdown initiates when the field amplitude exceeds approximately 2730.4 Oe, as illustrated in Fig. S5(a). At this field intensity, the skyrmion's guiding center traces a chaotic trajectory from the instant the field is applied up to its eventual breakdown. For the sake of clarity, we've approximated this value to 2730 Oe as the representative static magnetic field. By comparison, with a precise amplitude of 2730 Oe, the guiding center begins with chaotic motion and then transitions to stable trajectories with a gyration, as shown in Fig. S5(b). This gyration occurs at a frequency of 1.86 GHz, highlighting the altered eigenfrequency due to the skyrmion's deformation. The equilibrium state, marked by a gray dot, is located approximately at (1.31 nm, 1.14 nm). Interestingly, while the magnetic field was directed along the *y*-axis, the stabilization point is more skewed towards the *x*-axis.

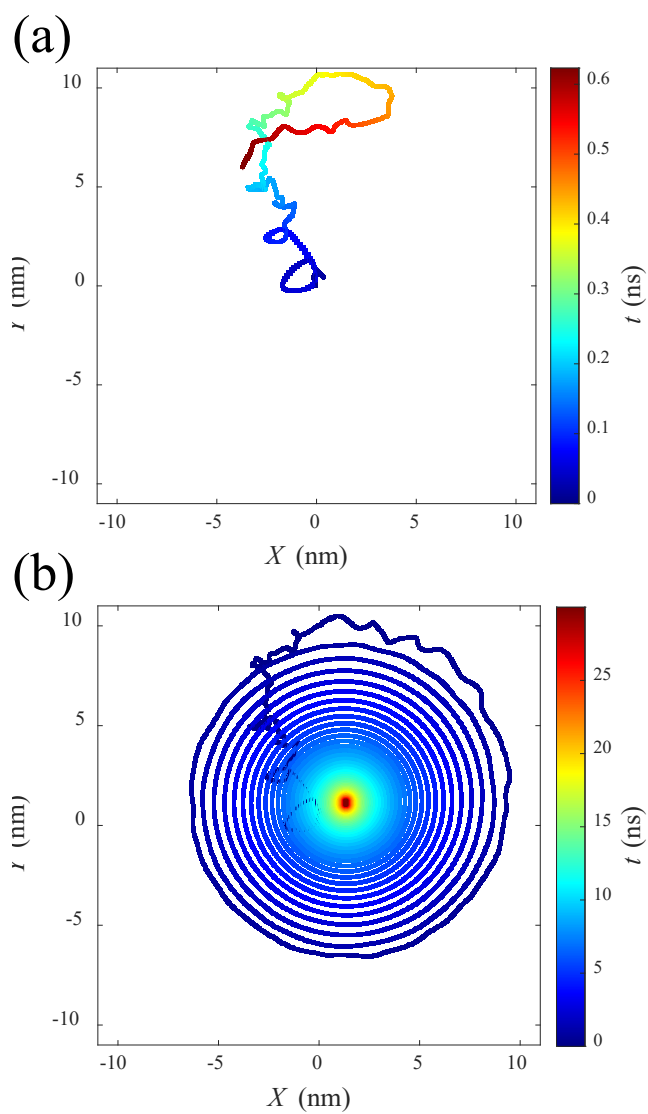

**Figure S5**. The trajectories of the guiding center under the influence of a static magnetic field applied along the y-axis. The field amplitude is 2730.4 Oe in (a) and 2730 Oe in (b). The progression of time is represented by the color index, and the gray dot indicates the equilibrium state.

**Appendix C: Sensitive dependence of Initial conditions**

| Initial condition | $<m_z>$ | q |
|---|---|---|
| 1 | 0.59717800 | -0.930778116463741 |
| 2 | 0.59716004 | -0.930784236318334 |
| 3 | 0.59715694 | -0.930785326955563 |
| 4 | 0.59712756 | -0.930795599997689 |
| 5 | 0.59711570 | -0.930799777066074 |

**Table S1**. Possible numerical values of ground states achieved from the relaxation of the same magnetic skyrmion are presented. Each state corresponds to a different initial condition. Extremely small differences between the values are denoted with an underline.

Regardless of the choice of differential equation solver, multiple numerical values for the skyrmion's ground state can be obtained during micromagnetic simulations. Among these, only five states were selected, as detailed in Table 1. The $<m_z> = M_z/M_s$ for every spin within the skyrmion exhibits differences up to the fifth decimal place. While these variations are too minimal to influence typical linear and nonlinear motions, they provide an excellent means to investigate SDI in the chaotic dynamics of the spin system. For instance, overlaying the time series of the *X*-signal from the five initial conditions, as depicted in Figure S6, offers a visualization of SDI. Even though all other conditions remain constant, the five distinct skyrmion ground states commence their movements following identical trajectories. However, their coherence is lost around 350 τ. The eventual breakdown of the skyrmion, inferred from the endpoint of each signal, occurs at markedly different times.

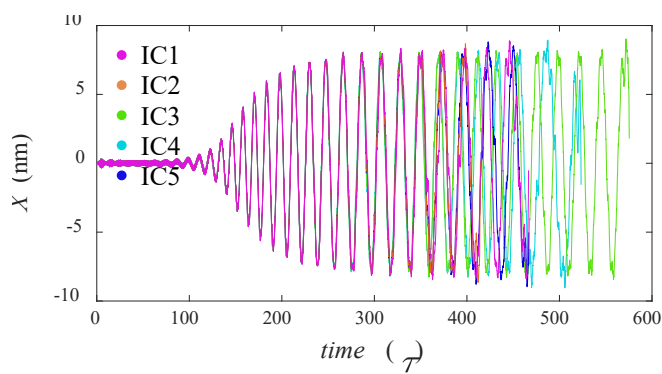

**Figure S6**. $X$-signal for five different initial conditions, labeled IC1 through IC5. The amplitude ($\alpha$) of the oscillating magnetic field is set to 0.16.

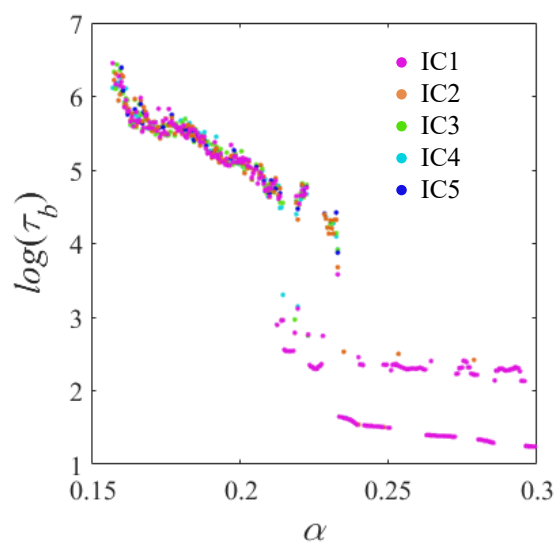

**Figure S7**. The breakdown time, $\tau_b$, of the skyrmion for different initial conditions on a logarithmic scale, labeled IC1 through IC5, is plotted with respect to the field amplitude, $\alpha$.

The SDI is particularly pronounced in the breakdown regime. The breakdown time, $\tau_b$, for the five different initial conditions have been plotted against varying field amplitude on a logarithmic scale, as depicted in Fig. S7. A skyrmion's breakdown in the chaotic regime can be interpreted as a manifestation of the self-criticality intrinsic to complex spin systems. It's widely accepted that self-criticality adheres to a power law, and in this instance, the breakdown time conforms to this law. Three distinct trend lines can be observed. In the first trend line, where α ranges from 0.157 to 0.233, and t spans from 3.5 to 6.5, the SDI effect is so pronounced that the skyrmion's breakdown time fluctuates significantly based on its initial condition. In contrast, the second (with a longer breakdown time) and third (with a shorter breakdown time) trends, where α lies between 0.233 and 0.3, exhibit a reduced SDI effect, suggesting that the chaos is, to some extent, mitigated.